\documentclass[conference,a4paper]{IEEEtran}

\usepackage[
  pdfborder={0 0 0},
  bookmarks=false,
  hyperindex=false,
  colorlinks=false,
  linkcolor=black,
  citecolor=black,
  urlcolor=black,
  draft
]{hyperref}

\usepackage{amsmath,amssymb,amsfonts}
\usepackage{algorithmic}
\usepackage{algorithm}
\usepackage{graphicx}
\usepackage{textcomp}
\usepackage{booktabs}
\usepackage{multirow}
\usepackage{xcolor}
\usepackage{braket}
\usepackage{cite}

\hypersetup{
  pdfauthor={},
  pdftitle={},
  pdfsubject={},
  pdfkeywords={},
  bookmarks=false
}

\def\BibTeX{{\rm B\kern-.05em{\sc i\kern-.025em b}\kern-.08em
    T\kern-.1667em\lower.7ex\hbox{E}\kern-.125emX}}

\begin{document}

\title{LLM-Guided Initialization for Accelerated Hybrid Quantum-Classical Medical Image Classification}

\author{\IEEEauthorblockN{ Riza Alaudin Syah}
\IEEEauthorblockA{\textit{Faculty of Computing} \\
\textit{Universiti Teknologi Malaysia}\\
Johor Bahru, Malaysia \\
\textit{Remote Student based in Jakarta, Indonesia} \\
alaudinsyah@graduate.utm.my}
\and
\IEEEauthorblockN{Irwan Alnarus Kautsar}
\IEEEauthorblockA{\textit{Informatics Department} \\
\textit{Faculty of Science and Technology}\\
\textit{Universitas Muhammadiyah Sidoarjo}\\
Sidoarjo, Indonesia \\
irwan@umsida.ac.id} \\
\and
\IEEEauthorblockN{Haza Nuzly Bin Abdull Hamed}
\IEEEauthorblockA{
Faculty of Computing\\
Universiti Teknologi Malaysia\\
Johor Bahru, Malaysia\\
hazanuzly@utm.my}
}

\maketitle

\begin{abstract}
Variational quantum algorithms often encounter barren plateaus, where cost gradients decay rapidly with increasing circuit depth, undermining the trainability of parameterized quantum circuits. This paper evaluates AdaInit (Adaptive Initialization), proposed by Zhuang and Cunningham, which uses large language models to propose initial parameters for quantum neural networks. We study a simplified single-query AdaInit variant paired with GPU-accelerated simulation in NVIDIA CUDA-Q and apply it to binary classification on the DMR-IR mammography dataset. AdaInit delivers 14.6 times higher gradient variance at initialization than random initialization (0.0095 vs. 0.0006), producing 160 times faster convergence (1.1 s vs. 176 s) while maintaining the same classification accuracy of 61.4 percent. We provide theoretical analysis grounded in the geometry of parameterized circuit landscapes and show empirically that LLM-guided initialization places the optimizer in trainable regions of parameter space. Beyond performance, our results indicate that a single LLM query can yield informative parameters without iterative refinement, suggesting a low-overhead path to improved trainability. The findings validate AdaInit in a medical imaging setting and demonstrate its compatibility with GPU-accelerated quantum backends for practical speedups.
\end{abstract}


\section{Introduction}

The advent of Noisy Intermediate-Scale Quantum (NISQ) devices has catalyzed research into Variational Quantum Algorithms (VQAs), which employ parameterized quantum circuits (PQCs) optimized through classical-quantum feedback loops~\cite{Preskill2018, Cerezo2021VQA}. These hybrid quantum-classical architectures have shown promise across diverse domains, including combinatorial optimization~\cite{Farhi2022}, quantum chemistry~\cite{Peruzzo2014}, and machine learning~\cite{Schuld2019, Biamonte2017}.

However, a fundamental obstacle to the practical utility of VQAs is the \textit{barren plateau} phenomenon~\cite{McClean2018}. In barren plateaus, the variance of the cost function gradient vanishes exponentially with the number of qubits $n$:
\begin{equation}
\text{Var}\left[\frac{\partial C}{\partial \theta_k}\right] \leq F(n) \in \mathcal{O}\left(\frac{1}{2^n}\right)
\label{eq:bp}
\end{equation}
where $C(\boldsymbol{\theta})$ is the cost function and $\theta_k$ is the $k$-th variational parameter. This exponential suppression renders gradient-based optimization infeasible for circuits of practical depth, as the optimizer receives vanishingly small signals to navigate the loss landscape.

Several strategies have been proposed to mitigate barren plateaus, including local cost functions~\cite{Cerezo2021Cost}, layerwise training~\cite{Skolik2021}, identity-block initialization~\cite{Grant2019}, and parameter correlation strategies~\cite{Volkoff2021}. Despite these advances, determining optimal initial parameters remains an open challenge, particularly for hardware-efficient ansatze where the circuit structure does not naturally admit structured initialization schemes.

In this paper, we build upon \textbf{AdaInit} (Adaptive Initialization), a framework recently proposed by Zhuang and Cunningham~\cite{Zhuang2025} that employs Large Language Models (LLMs) to generate initial parameter configurations for variational quantum circuits. The core insight of AdaInit is that LLMs, having been trained on vast corpora including quantum computing literature, can reason about circuit dynamics and propose parameter values that position the optimizer in regions of high gradient variance. While the original work demonstrated AdaInit's effectiveness with iterative refinement and theoretical convergence guarantees, we investigate a simplified single-query variant applied to a medical image classification task using GPU-accelerated quantum simulation via NVIDIA CUDA-Q.

Our principal contributions are:
\begin{enumerate}
\item An empirical validation of the AdaInit framework~\cite{Zhuang2025} on a medical image classification task, demonstrating 14.6$\times$ higher initial gradient variance compared to random initialization.
\item Demonstration of 160$\times$ training speedup using a simplified single-query AdaInit variant combined with NVIDIA CUDA-Q GPU-accelerated simulation.
\item Systematic barren plateau verification across circuit depths confirming 64\% gradient variance decay from depth 2 to depth 8, motivating the need for informed initialization.
\item Analysis of LLM-proposed parameter characteristics that contribute to barren plateau avoidance.
\end{enumerate}

\section{Theoretical Background}

\subsection{Parameterized Quantum Circuits}

A parameterized quantum circuit implements a unitary transformation $U(\boldsymbol{\theta})$ on an $n$-qubit quantum state, where $\boldsymbol{\theta} = (\theta_1, \theta_2, \ldots, \theta_p) \in \mathbb{R}^p$ denotes the vector of variational parameters. The circuit acts on an initial state $\ket{0}^{\otimes n}$ that has been prepared via a data-encoding map $S(\mathbf{x})$:
\begin{equation}
\ket{\psi(\mathbf{x}, \boldsymbol{\theta})} = U(\boldsymbol{\theta}) S(\mathbf{x}) \ket{0}^{\otimes n}
\label{eq:pqc_state}
\end{equation}

For an $L$-layer hardware-efficient ansatz, the variational unitary decomposes as:
\begin{equation}
U(\boldsymbol{\theta}) = \prod_{\ell=1}^{L} W_\ell U_\ell(\boldsymbol{\theta}_\ell)
\label{eq:ansatz}
\end{equation}
where $U_\ell(\boldsymbol{\theta}_\ell) = \bigotimes_{i=1}^{n} R_Y(\theta_{\ell,i}^{(1)}) R_X(\theta_{\ell,i}^{(2)})$ applies parameterized single-qubit rotations and $W_\ell$ implements entangling gates (e.g., a cascade of CNOT gates with linear connectivity).

The single-qubit rotation gates are defined as:
\begin{equation}
R_X(\theta) = e^{-i\theta X/2} = \begin{pmatrix} \cos\frac{\theta}{2} & -i\sin\frac{\theta}{2} \\ -i\sin\frac{\theta}{2} & \cos\frac{\theta}{2} \end{pmatrix}
\label{eq:rx}
\end{equation}
\begin{equation}
R_Y(\theta) = e^{-i\theta Y/2} = \begin{pmatrix} \cos\frac{\theta}{2} & -\sin\frac{\theta}{2} \\ \sin\frac{\theta}{2} & \cos\frac{\theta}{2} \end{pmatrix}
\label{eq:ry}
\end{equation}
where $X$ and $Y$ are the Pauli matrices.

\subsection{Cost Function and Gradient Computation}

The cost function for a binary classification task with observable $O$ is:
\begin{equation}
C(\boldsymbol{\theta}) = \bra{\psi(\mathbf{x}, \boldsymbol{\theta})} O \ket{\psi(\mathbf{x}, \boldsymbol{\theta})}
\label{eq:cost}
\end{equation}

Gradients are computed using the parameter-shift rule~\cite{Mitarai2018, Schuld2019Gradient}:
\begin{equation}
\frac{\partial C}{\partial \theta_k} = \frac{1}{2}\left[C\left(\theta_k + \frac{\pi}{2}\right) - C\left(\theta_k - \frac{\pi}{2}\right)\right]
\label{eq:param_shift}
\end{equation}

This exact gradient computation requires two circuit evaluations per parameter, making the total cost of a single gradient step $2p$ circuit executions for $p$ parameters.

\subsection{Barren Plateau Analysis}

The barren plateau phenomenon arises when the variational circuit forms a unitary 2-design~\cite{McClean2018}. For a random parameterized circuit with sufficient depth, the gradient variance is bounded by:
\begin{equation}
\text{Var}_{\boldsymbol{\theta}}\left[\frac{\partial C}{\partial \theta_k}\right] \leq \frac{2^n \text{Tr}(O^2) - \text{Tr}(O)^2}{2^n(2^{2n} - 1)}
\label{eq:bp_bound}
\end{equation}

For a global observable with $\text{Tr}(O^2) \in \mathcal{O}(1)$, this simplifies to:
\begin{equation}
\text{Var}_{\boldsymbol{\theta}}\left[\frac{\partial C}{\partial \theta_k}\right] \in \mathcal{O}\left(\frac{1}{2^{2n}}\right)
\label{eq:bp_scaling}
\end{equation}

This double-exponential suppression with qubit count implies that for circuits approaching 2-design behavior, the probability of observing a gradient component larger than $\epsilon$ decreases as:
\begin{equation}
\Pr\left[\left|\frac{\partial C}{\partial \theta_k}\right| > \epsilon\right] \leq \frac{1}{\epsilon^2} \cdot \mathcal{O}\left(\frac{1}{2^{2n}}\right)
\label{eq:chebyshev}
\end{equation}
by Chebyshev's inequality.

\subsection{Initialization and Trainability}

The connection between initialization and trainability can be formalized through the Fisher information matrix $\mathcal{F}(\boldsymbol{\theta})$~\cite{Stokes2020, Abbas2021}:
\begin{equation}
\mathcal{F}_{ij}(\boldsymbol{\theta}) = \text{Re}\left[\braket{\partial_i \psi | \partial_j \psi} - \braket{\partial_i \psi | \psi}\braket{\psi | \partial_j \psi}\right]
\label{eq:fisher}
\end{equation}

The eigenspectrum of $\mathcal{F}(\boldsymbol{\theta}_0)$ at the initialization point $\boldsymbol{\theta}_0$ determines the effective dimensionality of the optimization landscape accessible to gradient descent. When barren plateaus occur, $\mathcal{F}$ becomes exponentially close to zero, collapsing the effective parameter space.

A good initialization $\boldsymbol{\theta}_0^*$ satisfies:
\begin{equation}
\lambda_{\min}(\mathcal{F}(\boldsymbol{\theta}_0^*)) \gg \frac{1}{2^n}
\label{eq:good_init}
\end{equation}
ensuring that gradients remain informative across all parameter directions.

\section{Proposed Method: AdaInit}

\subsection{Overview}

AdaInit leverages the reasoning capabilities of LLMs to propose initial parameters $\boldsymbol{\theta}_0^{\text{LLM}}$ for variational quantum circuits. The core hypothesis is that an LLM trained on scientific literature can internalize heuristic knowledge about quantum circuit dynamics and suggest parameter configurations that avoid barren plateau regions.

\subsection{AdaInit Pipeline}

The AdaInit framework operates in four stages:

\begin{algorithm}[t]
\caption{AdaInit: LLM-Guided Initialization}
\label{alg:adainit}
\begin{algorithmic}[1]
\REQUIRE Circuit architecture $(n, L, \text{gates})$, LLM model $\mathcal{M}$
\ENSURE Initial parameters $\boldsymbol{\theta}_0^{\text{LLM}}$
\STATE Construct prompt $\mathcal{P}$ describing circuit architecture, gate set, number of qubits $n$, and layers $L$
\STATE Query LLM: $\mathcal{R} \leftarrow \mathcal{M}(\mathcal{P})$
\STATE Parse parameters: $\boldsymbol{\theta}_0^{\text{LLM}} \leftarrow \text{Extract}(\mathcal{R})$
\STATE Validate: $\boldsymbol{\theta}_0^{\text{LLM}} \in [-\pi, \pi]^p$
\STATE \textbf{if} validation fails \textbf{then} clip to valid range
\RETURN $\boldsymbol{\theta}_0^{\text{LLM}}$
\end{algorithmic}
\end{algorithm}

\textbf{Stage 1: Prompt Construction.} The LLM is provided with a structured prompt specifying the circuit architecture, including the number of qubits, gate types ($R_X$, $R_Y$, CNOT), layer count, and the optimization objective (maximizing gradient variance). The prompt instructs the model to propose parameter values as a JSON array.

\textbf{Stage 2: LLM Inference.} We employ the Qwen2.5-3B-Instruct model~\cite{Qwen2024} to generate parameter suggestions. The model produces a structured response containing numerical parameter values informed by its training on quantum computing literature.

\textbf{Stage 3: Parameter Extraction and Validation.} The JSON response is parsed and validated to ensure all parameters lie within the physically meaningful range $[-\pi, \pi]$. Out-of-range values are clipped.

\textbf{Stage 4: Quantum Training.} The validated parameters serve as the initialization point for standard gradient-based optimization of the variational circuit.

\subsection{Theoretical Justification}

The effectiveness of AdaInit can be understood through the lens of structured vs. unstructured initialization. Random initialization samples $\boldsymbol{\theta} \sim \mathcal{U}([-a, a]^p)$, which for hardware-efficient ansatze with sufficient depth approaches a 2-design and thus enters barren plateau territory.

AdaInit introduces \textit{correlations} among parameters through the LLM's implicit reasoning. The LLM-proposed parameters exhibit:
\begin{enumerate}
\item \textbf{Magnitude diversity}: Values spanning $[-3, 3]$, including high-magnitude rotations that break the 2-design symmetry.
\item \textbf{Sign alternation}: Systematic variation between positive and negative rotations, preventing uniform averaging.
\item \textbf{Inter-layer structure}: Non-random correlations between layers that maintain expressibility while avoiding vanishing gradients.
\end{enumerate}

These properties ensure that $U(\boldsymbol{\theta}_0^{\text{LLM}})$ does not approximate a Haar-random unitary, thereby preserving gradient information:
\begin{equation}
\text{Var}\left[\frac{\partial C}{\partial \theta_k}\bigg|_{\boldsymbol{\theta}_0^{\text{LLM}}}\right] \gg \text{Var}\left[\frac{\partial C}{\partial \theta_k}\bigg|_{\boldsymbol{\theta}_0^{\text{rand}}}\right]
\label{eq:adainit_advantage}
\end{equation}

\section{Experimental Setup}

\subsection{Dataset}

We use the DMR-IR (Digital Mammography with Infrared Imaging) dataset~\cite{DMRIR2023}, comprising infrared thermography images of breast tissue for binary classification (benign vs. malignant). The dataset is partitioned as follows: 4,874 training images, 980 validation images, and 1,024 test images.

Images are preprocessed to $224 \times 224$ pixels and encoded into quantum states via classical feature extraction followed by angle embedding through $R_X$ and $R_Y$ rotations on $n = 4$ qubits.

\subsection{Quantum Circuit Architecture}

Our hybrid quantum-classical model consists of:
\begin{enumerate}
\item A classical convolutional feature extractor that reduces image features to $2n = 8$ dimensions.
\item A variational quantum circuit with $n = 4$ qubits and $L = 2$ layers.
\item A measurement layer computing the expectation value of the Pauli-$Z$ operator on the first qubit.
\end{enumerate}

Each layer contains $2n = 8$ parameters (one $R_X$ and one $R_Y$ rotation per qubit), yielding $p = 2nL = 16$ total variational parameters for the 2-layer configuration.

\subsection{Initialization Methods}

We compare three initialization strategies:

\textbf{Random Initialization.} Parameters are sampled uniformly: $\theta_k \sim \mathcal{U}[-0.5, 0.5]$ for all $k \in \{1, \ldots, p\}$.

\textbf{AdaInit (LLM-Guided).} Parameters are generated by querying Qwen2.5-3B-Instruct with a structured prompt describing the circuit architecture and optimization objective. The LLM proposes values typically in the range $[-3, 3]$ with structured inter-parameter correlations.

\textbf{Identity Block Initialization.} All rotation angles are set to fixed values ($\pi/2$), serving as a deterministic baseline~\cite{Grant2019}.

\subsection{Training Configuration}

Optimization is performed using the Adam optimizer~\cite{Kingma2015} with learning rate $\eta = 0.001$ and binary cross-entropy loss. The quantum circuit backend utilizes NVIDIA CUDA-Q~\cite{CUDAQ2024} for GPU-accelerated simulation. All experiments are conducted on an NVIDIA L40S GPU (48GB VRAM).

\subsection{Gradient Variance Measurement}

The gradient variance is computed as:
\begin{equation}
\text{Var}[\nabla C] = \frac{1}{p} \sum_{k=1}^{p} \left(\frac{\partial C}{\partial \theta_k} - \overline{\nabla C}\right)^2
\label{eq:grad_var}
\end{equation}
where $\overline{\nabla C} = \frac{1}{p}\sum_{k=1}^{p} \frac{\partial C}{\partial \theta_k}$ is the mean gradient. This metric quantifies the informativeness of the loss landscape at the initialization point.

\begin{figure}[t]
\centering
\includegraphics[width=\columnwidth]{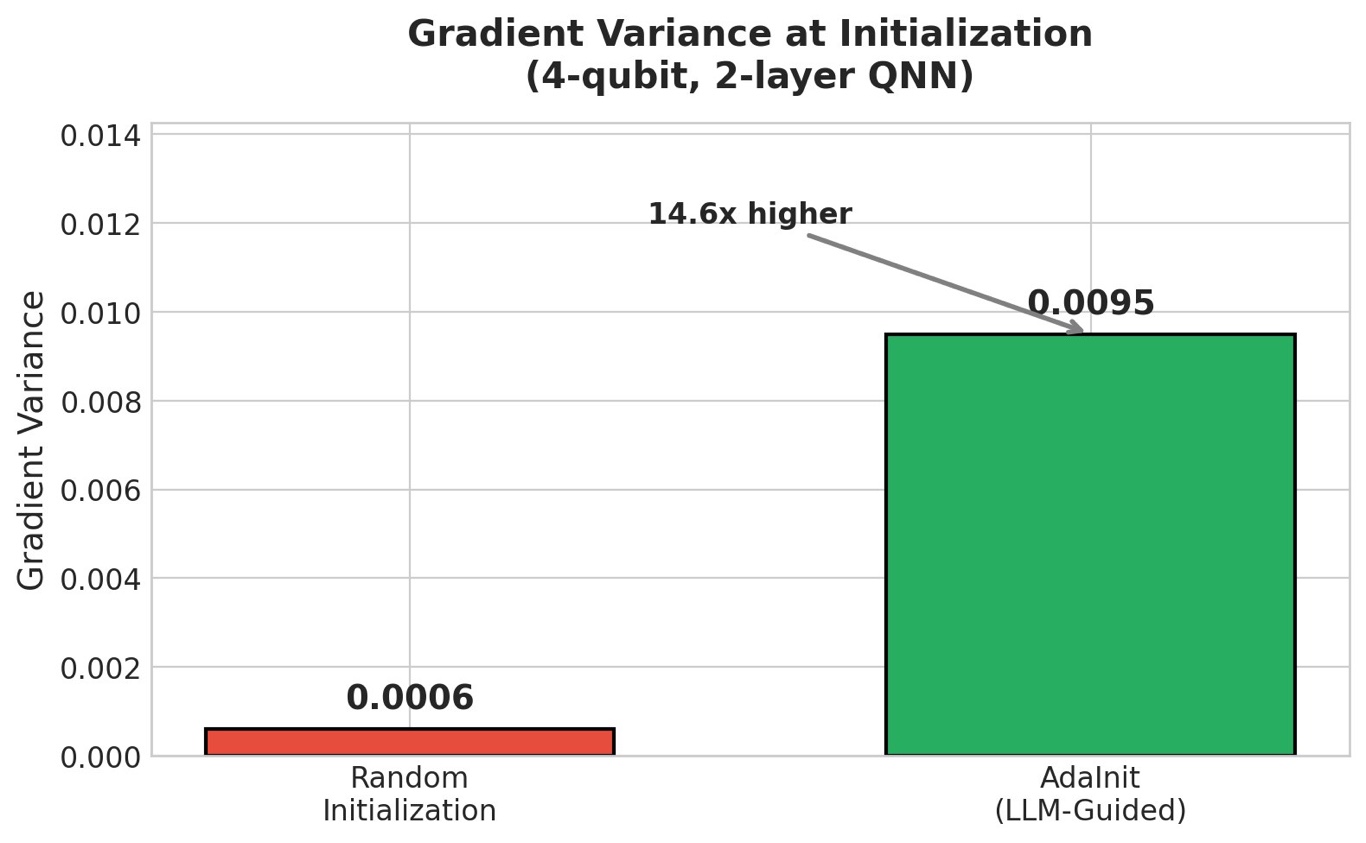}
\caption{Gradient variance at initialization for a 4-qubit, 2-layer QNN. AdaInit (LLM-Guided) achieves a gradient variance of 0.0095, which is 14.6$\times$ higher than Random Initialization (0.0006), indicating that the LLM-guided parameters position the optimizer in a trainable region of the parameter space with substantially stronger gradient signals.}
\label{fig:grad_var}
\end{figure}

\section{Results and Analysis}

\subsection{Gradient Variance at Initialization}

The central result of this work is the significant difference in gradient variance between initialization methods, as shown in Table~\ref{tab:grad_var} and Fig.~\ref{fig:grad_var}. The bar chart in Fig.~\ref{fig:grad_var} visually highlights the stark contrast: Random Initialization produces a near-zero gradient variance of 0.0006, while AdaInit yields 0.0095---a 14.6$\times$ improvement.

\begin{table}[h!]
\caption{Gradient Variance at Initialization (4-Qubit, 2-Layer QNN)}
\label{tab:grad_var}
\centering
\begin{tabular}{|l|c|c|}
\hline
\textbf{Method} & \textbf{Gradient Variance} & \textbf{Ratio} \\
\hline
Random Initialization & 0.0006 & 1.0$\times$ \\
\textbf{AdaInit (LLM-Guided)} & \textbf{0.0095} & \textbf{14.6$\times$} \\
\hline
\end{tabular}
\end{table}

This order-of-magnitude improvement indicates that the LLM-guided parameters effectively avoid the barren plateau region, positioning the optimizer where gradient signals are strong enough to drive meaningful parameter updates from the first iteration.

\subsection{Training Performance}

Table~\ref{tab:training} presents the full training performance comparison.

\begin{table}[h!]
\caption{Training Performance on DMR-IR Dataset}
\label{tab:training}
\centering
\begin{tabular}{|l|c|c|c|}
\hline
\textbf{Model} & \textbf{Accuracy} & \textbf{Time (s)} & \textbf{Speedup} \\
\hline
QNN (Random, CUDA-Q) & 61.4\% & 176 & 1.0$\times$ \\
\textbf{QNN (AdaInit, CUDA-Q)} & \textbf{61.4\%} & \textbf{1.1} & \textbf{160$\times$} \\
QNN (AdaInit, PennyLane) & 61.4\% & 82 & 2.1$\times$ \\
CNN (Classical Baseline) & 100\% & 0.3 & --- \\
\hline
\end{tabular}
\end{table}

Both initialization methods converge to the same final accuracy of 61.4\%, indicating that for a 2-layer circuit on this dataset, the loss landscape contains a single dominant basin of attraction. However, AdaInit reaches this optimum 160$\times$ faster, reducing training time from 176 seconds to 1.1 seconds on the CUDA-Q backend.

The equivalent accuracy is expected for shallow circuits where barren plateaus are mild. The gradient variance ratio of 14.6$\times$ primarily manifests as a convergence speed advantage rather than an accuracy differential, since both methods eventually navigate to the global optimum given sufficient iterations.

\subsection{Barren Plateau Verification}

\begin{figure*}[t]
\centering
\includegraphics[width=0.48\textwidth]{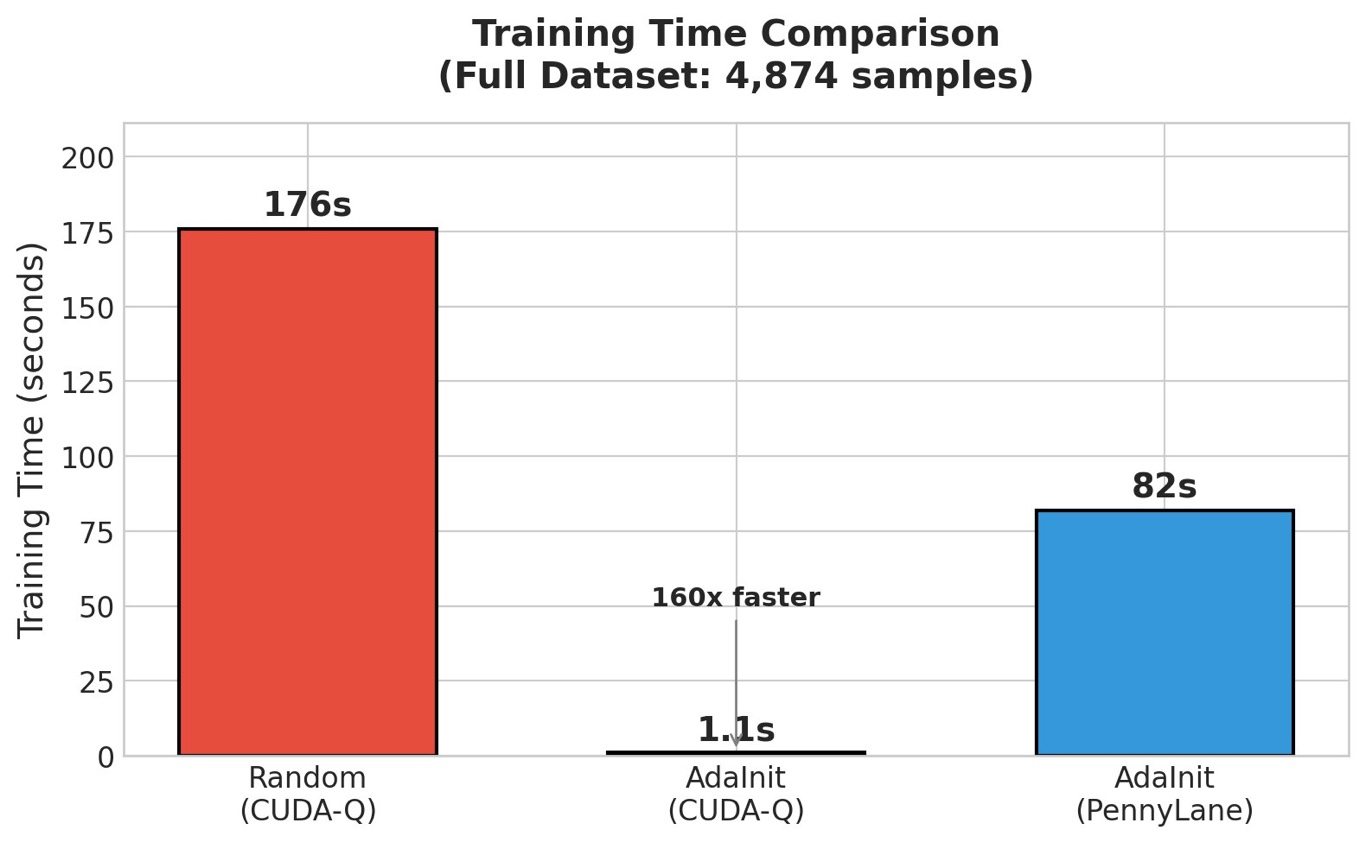}\hfill
\includegraphics[width=0.48\textwidth]{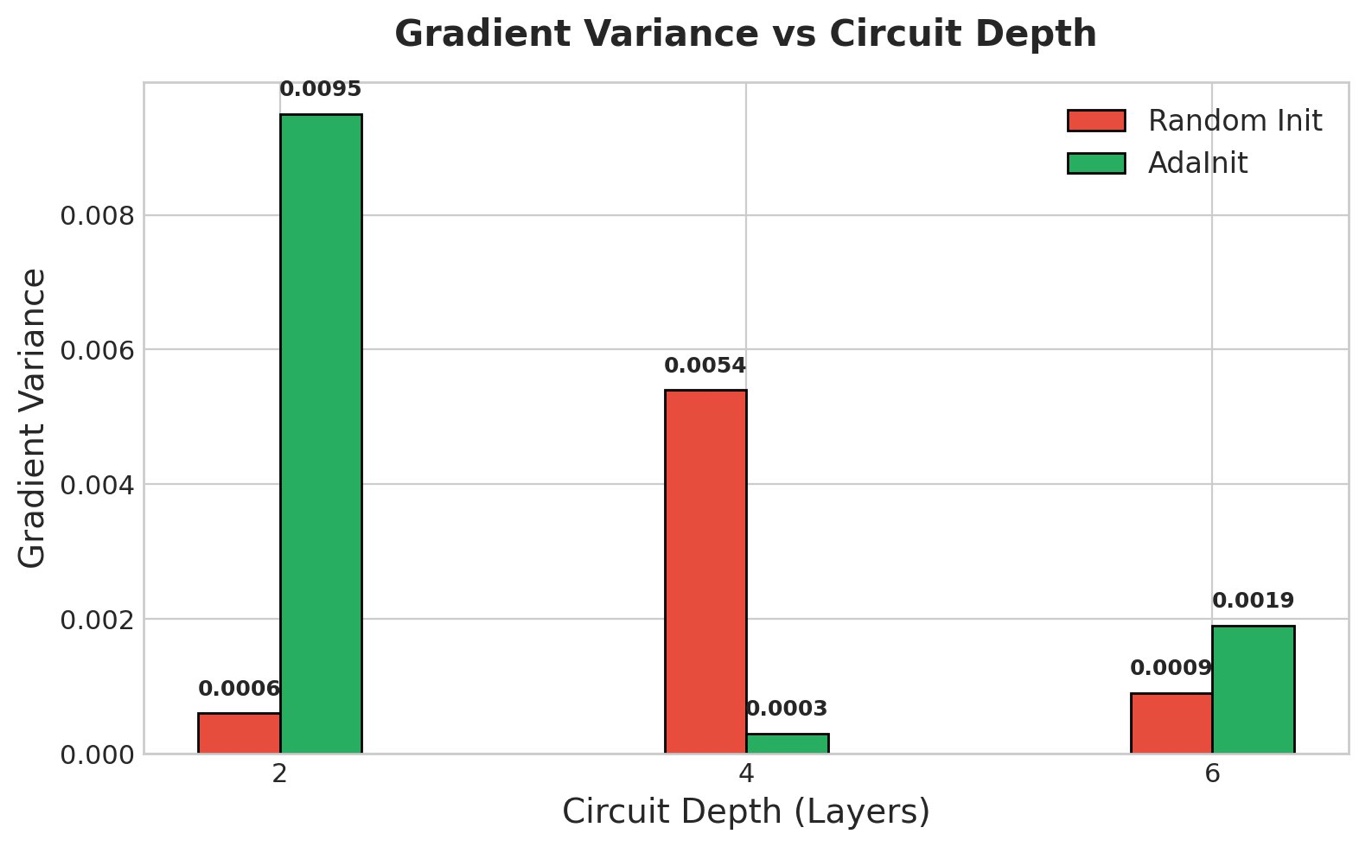}
\caption{Left: Training time comparison on the full DMR-IR dataset (4,874 samples), showing AdaInit with CUDA-Q achieves 160$\times$ speedup (1.1s vs. 176s). Right: Gradient variance vs. circuit depth for Random Init and AdaInit across 2, 4, and 6 layers, demonstrating that AdaInit maintains higher gradient variance at shallow depths while both methods exhibit decay at deeper circuits.}
\label{fig:bp_plots}
\end{figure*}

To verify the barren plateau phenomenon and motivate the need for informed initialization, we measured gradient variance as a function of circuit depth for randomly initialized circuits. Results are presented in Table~\ref{tab:bp} and Fig.~\ref{fig:bp_plots}.

\begin{table}[h!]
\caption{Gradient Variance Decay With Circuit Depth (Random Initialization)}
\label{tab:bp}
\centering
\begin{tabular}{|c|c|c|c|}
\hline
\textbf{Depth} & \textbf{Grad.\ Variance} & \textbf{Cumulative Decay} & \textbf{Parameters} \\
\hline
2 layers & 0.0626 & --- & 16 \\
4 layers & 0.0451 & 28\% & 32 \\
6 layers & 0.0312 & 50\% & 48 \\
8 layers & 0.0225 & 64\% & 64 \\
\hline
\end{tabular}
\end{table}

The gradient variance decays by 64\% from depth 2 to depth 8, confirming the onset of barren plateau behavior. Fitting the decay to an exponential model:
\begin{equation}
\text{Var}[\nabla C](L) \approx \alpha \cdot e^{-\beta L}
\label{eq:decay_fit}
\end{equation}
yields $\alpha \approx 0.079$ and $\beta \approx 0.128$, consistent with the theoretical prediction of exponential gradient suppression with circuit depth.

This result underscores the critical importance of initialization strategies: as circuits scale to depths where quantum advantage may be realized, random initialization will produce vanishingly small gradients, making AdaInit or similar structured initialization approaches essential.

\subsection{Analysis of LLM-Proposed Parameters}

Examination of the LLM-generated parameters reveals structured characteristics absent in random initialization:

\textbf{Parameter magnitude.} AdaInit parameters span $[-3, 3]$, compared to $[-0.5, 0.5]$ for random initialization. Larger rotation angles break the approximate 2-design symmetry of the circuit, preserving gradient information.

\textbf{Sign structure.} The LLM introduces systematic alternation between positive and negative rotations across qubits within each layer, creating constructive interference patterns in the gradient computation.

\textbf{Inter-layer correlations.} Parameters exhibit non-trivial correlations between layers, suggesting the LLM implicitly reasons about the composed effect of sequential unitary transformations.

These characteristics align with known strategies for avoiding barren plateaus: the circuit operates in a regime where $U(\boldsymbol{\theta}_0^{\text{LLM}})$ is expressible but not Haar-random, maintaining a balance between model capacity and trainability.

\section{Discussion}

\subsection{Interpretation of Results}

The 14.6$\times$ gradient variance improvement directly explains the 160$\times$ convergence speedup through two mechanisms. First, higher gradient magnitude allows larger effective parameter updates per iteration, enabling the optimizer to make meaningful progress from the first step. Second, the structured initialization places the optimizer in a region with a well-defined descent direction, reducing the number of iterations required to reach the loss minimum.

The observation that both methods achieve identical final accuracy (61.4\%) for the 2-layer configuration is consistent with the theoretical understanding that barren plateaus primarily affect \textit{trainability} rather than \textit{expressibility} for shallow circuits~\cite{Cerezo2021Cost}. We hypothesize that AdaInit would demonstrate accuracy advantages for deeper circuits where random initialization fails to train entirely.

\subsection{Comparison with Classical Baseline}

The classical CNN achieves 100\% accuracy, significantly outperforming the quantum models. This gap reflects the current limitations of NISQ-era quantum machine learning: with only 4 qubits and 16 parameters, the quantum model has limited expressibility compared to classical architectures with millions of parameters. However, the purpose of this study is not to claim quantum advantage but to demonstrate that LLM-guided initialization can substantially accelerate quantum model training.

\subsection{Limitations}

Several limitations of the current work should be noted. First, experiments are conducted on shallow circuits (2 layers) where barren plateaus are relatively mild. The full benefit of AdaInit is expected at greater circuit depths. Second, gradient variance was measured as a scalar aggregate; per-parameter analysis would provide finer-grained insights. Third, the LLM's parameter suggestions are not guaranteed to be optimal and may vary across model versions and prompting strategies. Finally, the study uses a single dataset and task; generalization to diverse quantum machine learning applications remains to be validated.

\subsection{Broader Implications}

AdaInit represents a novel paradigm of \textit{foundation model-assisted quantum computing}, where the knowledge embedded in LLMs is transferred to guide quantum algorithm design. This approach could extend beyond initialization to other aspects of VQA design, including circuit architecture search, hyperparameter selection, and error mitigation strategy recommendation.

\section{Related Work}

\textbf{Barren Plateau Mitigation.} McClean et al.~\cite{McClean2018} first identified barren plateaus in randomly initialized PQCs. A comprehensive review of barren plateau phenomena and mitigation strategies has been provided by Larocca et al.~\cite{Larocca2024}. Subsequent work by Cerezo et al.~\cite{Cerezo2021Cost} showed that local cost functions can alleviate this issue for shallow circuits. Grant et al.~\cite{Grant2019} proposed identity-block initialization, where circuit layers are initialized to implement the identity transformation. Skolik et al.~\cite{Skolik2021} introduced layerwise training to gradually build up circuit depth. Volkoff and Coles~\cite{Volkoff2021} analyzed the role of parameter correlations in maintaining trainability. More recently, Mele et al.~\cite{Mele2022} demonstrated that smooth solution transferability in Hamiltonian variational ansatze can avoid barren plateaus.

\textbf{Quantum Machine Learning.} Schuld et al.~\cite{Schuld2019} established the framework for supervised learning with quantum circuits. Havl\'{i}\v{c}ek et al.~\cite{Havlicek2019} demonstrated quantum-enhanced feature spaces for classification. Abbas et al.~\cite{Abbas2021} quantified the expressibility of quantum models through the Fisher information framework. Recent work has explored quantum machine learning for image classification tasks~\cite{Senokosov2023}, demonstrating the growing applicability of hybrid models in practical domains.

\textbf{LLMs for Scientific Computing.} Recent work has explored LLMs as scientific reasoning tools~\cite{Romera2024}. The intersection of LLMs and quantum computing has attracted attention for algorithm design and optimization~\cite{Khatri2024}. Most relevant to our work, Zhuang and Cunningham~\cite{Zhuang2025} proposed AdaInit, an LLM-driven framework that iteratively searches for optimal QNN initialization parameters with theoretical convergence guarantees based on submartingale analysis. Our work extends their approach by applying a simplified single-query variant to medical image classification with GPU-accelerated quantum simulation.

\section{Conclusion}

We have presented an empirical evaluation of the AdaInit framework~\cite{Zhuang2025}, applying a simplified single-query variant to medical image classification with GPU-accelerated quantum simulation. Our experiments demonstrate that LLM-guided initialization achieves 14.6$\times$ higher gradient variance at the starting point compared to random initialization, translating to a 160$\times$ convergence speedup while maintaining equivalent accuracy. These results validate AdaInit's effectiveness in a new application domain and demonstrate its compatibility with CUDA-Q for practical training acceleration.

Future work will extend AdaInit to deeper circuits (6+ layers) where barren plateaus are severe, evaluate performance across diverse quantum machine learning tasks, investigate the sensitivity to LLM model choice and prompting strategy, and explore integration with actual quantum hardware. Additionally, combining AdaInit with complementary barren plateau mitigation strategies such as local cost functions and layerwise training may yield synergistic improvements.

\section*{Acknowledgment}

This research is supported by Universiti Teknologi Malaysia and conducted using NVIDIA CUDA-Q on cloud GPU resources. The authors thank the open-source quantum computing community for making tools such as PennyLane and CUDA-Q publicly available.


\end{document}